\documentclass[3p,times]{elsarticle}

\usepackage{ecrc}
\usepackage{natbib}
\volume{00}

\firstpage{1}

\journalname{Materials Chemistry and Physics}

\runauth{ V. Atanasov et. al. }

\jid{matchemphys}

\jnltitlelogo{Materials Chemistry and Physics}

\CopyrightLine{2015}{Published by Elsevier Ltd.}

\begin{document}

\begin{frontmatter}



\dochead{}
%% e.g. \dochead{17th International Conference on Dynamical Processes in Excited States of Solids}

\title{Two Dimensional Polymerization of Graphene Oxide: Bottom-up Approach}


\author[label1]{Victor Atanasov\footnote{ {\it Tel.:} +359 894 431034}}
\address[label1]{Sofia University, Faculty of Physics, 5 boul. J. Bourchier, 1164 Sofia, Bulgaria}
\ead{vatanaso@phys.uni-sofia.bg}
\ead[url]{http://phys.uni-sofia.bg/~vatanaso/}

\author[label1]{Stoyan Russev} 
%\address{Sofia University, Faculty of Physics, 5 boul. J. Bourchier, 1164 Sofia, Bulgaria}
%\email{scr@phys.uni-sofia.bg}

\author[label2]{Lyudmil Lyutov} 
\address[label2]{Sofia University, Faculty of Chemistry, 1 boul. J. Bourchier, 1164 Sofia, Bulgaria}
%\email{nhll@chem.uni-sofia.bg}

\author[label2]{Yulian Zagranyarski}
%\address{Sofia University, Faculty of Chemistry, 1 boul. J. Bourchier, 1164 Sofia, Bulgaria}
%\email{ohjz@chem.uni-sofia.bg}

\author[label2]{Iglika Dimitrova}
%\address{Sofia University, Faculty of Chemistry, 1 boul. J. Bourchier, 1164 Sofia, Bulgaria}
%\email{igli88@abv.bg}

\author[label3]{Georgy Avdeev} 
\address[label3]{Bulgarian Academy of Sciences, Acad. G. Bonchev, Str. , Block 11, Sofia 1113, Bulgaria}
%\email{g_avdeev@abv.bg}

\author[label3]{Ivalina Avramova} 
%\address{Bulgarian Academy of Sciences, Acad. G. Bonchev, Str. , Block 11, Sofia 1113, Bulgaria}
%\email{iva@svr.igic.bas.bg}

\author[label1]{Evgenia Vulcheva}
%\address{Sofia University, Faculty of Physics, 5 boul. J. Bourchier, 1164 Sofia, Bulgaria}
%\email{epv@phys.uni-sofia.bg}

\author[label1]{Kiril Kirilov}
%\address{Sofia University, Faculty of Physics, 5 boul. J. Bourchier, 1164 Sofia, Bulgaria}
%\email{kirilowk@phys.uni-sofia.bg}

\author[label1]{Atanas Tzonev}
%\address{Sofia University, Faculty of Physics, 5 boul. J. Bourchier, 1164 Sofia, Bulgaria}
%\email{atanas_tzonev@abv.bg}

\author[label1]{Miroslav Abrashev}  
%\address{Sofia University, Faculty of Physics, 5 boul. J. Bourchier, 1164 Sofia, Bulgaria}
%\email{mvabr@phys.uni-sofia.bg}

\author[label1]{Gichka Tsutsumanova}
%\address{Sofia University, Faculty of Physics, 5 boul. J. Bourchier, 1164 Sofia, Bulgaria}
%\email{ggt@phys.uni-sofia.bg}

\begin{abstract}
We demonstrate a bottom-up synthesis of structures similar to graphene oxide via a two dimensional polymerization. Experimental evidence and discussion are conveyed as well as a general framework for this two dimensional polymerization. The proposed morphologies and lattice structures of these graphene oxides are derived from aldol condensation of alternating three nucleophilic and three electrophilic centers of benzenetriol.
\end{abstract}

\begin{keyword}

chemical synthesis, polymers, chemical techniques, monolayers, electron diffraction




\end{keyword}

\end{frontmatter}

\section{Introduction}
The synergy between unmatched electrical, optical and mechanical properties of graphene and graphene oxide (GO) has resulted in vigorous research into methods for their large-scale production \citep{gr1, gr2, gr3}. 

Sheets of GO with atomic thickness have established themselves as a new carbon-based nanoscale material with an optical band gap of 1.7 eV to 2.1 eV \citep{go_gap} and soluble in water and other solvents which allows it to be spray or spin coated\citep{go_water, go_micro}. Controlled oxidation provides tunability of its electronic and mechanical properties up to the point of its turning into the semi-metallic graphene upon complete removal of the C-O bonds. Therefore, we assume the synthesis of GO via a two dimensional polymerization a successful attempt at bottom-up synthesis of graphene.

In general, chemical oxidation methods such as  Brodie's \citep{brodie}, Staudenmaier's \citep{st}, Hummers' \citep{hm} or a variation of these\citep{tour}, produce GO by introducing functional groups in between the layers forming graphite and they peel off. However, it represents a top-down approach.

In this communication we suggest a bottom-up approach towards GO synthesis. We start from a simple monomer and through a tailored two dimensional polymerization arrive at GO platelets. 

\begin{figure}[h]
\centering
  \includegraphics[height=6.5cm]{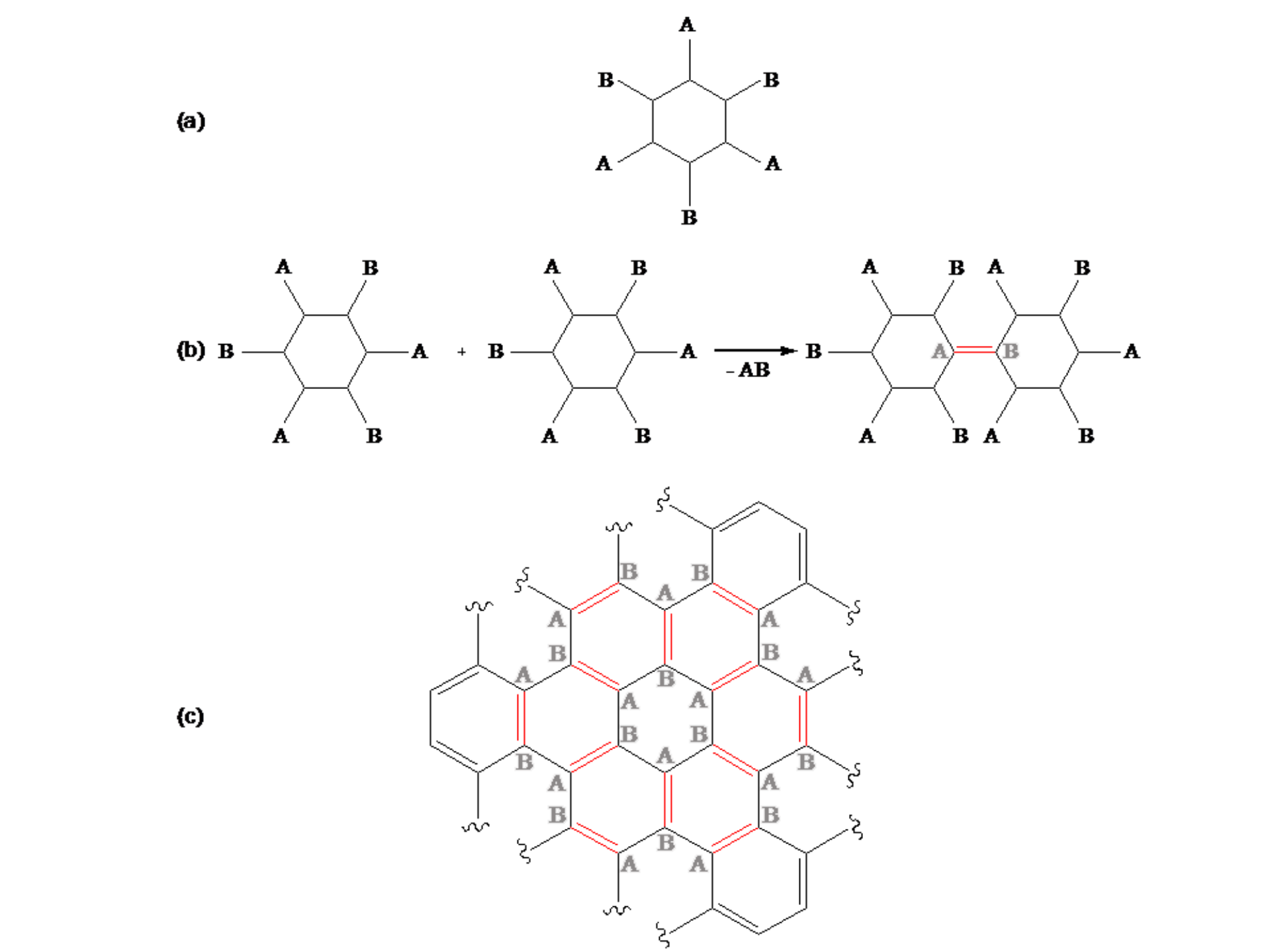}
  \caption{(a.) Monomer unit; (b.) Polymerization rule; (c.) Upon polymerization the bonds formed between the benzene rings span a graphene plane.  \label{fig1}}
\end{figure}

Consider the monomer unit in Fig \ref{fig1}a. This unit contains alternating functional groups A and B attached to a benzene or cyclohexane carbon ring. Imposing a polymerization rule that depends on the chemical reaction between A and B we can produce the step depicted in Fig \ref{fig1}b. The newly formed bond (depicted in color) appears in an arbitrary position in the ring with equal probability for all positions. The letters in the ring denote the positions functional groups had prior to the reaction. Given this rule, one is tempted to assume that the A and B functional groups above and below the bond would also react. However, they are attached with covalent bonds to an sp$^{n}$ hybridized carbon atom and these bonds are rigid. Such a reaction would require their bending and this is energy prohibitive. More probable is the reaction with an additional monomer unit. When this occurs we arrive at the lattice represented in Fig \ref{fig1}c. The hexagonal symmetry is present in this newly formed structure. The bonds formed during the polymerization are depicted in color. The key to the understanding of the formation of this lattice is the observation that the intermediate (to the monomer rings) hexagons are spanned by three newly formed bonds.  If one traces the central monomer hexagon and the surrounding it six monomer hexagons, one is convinced that this is the only possible configuration the original monomer units undergone two dimensional polymerization can arrange in space.

\begin{figure}[h]
\centering
  \includegraphics[height=6.5cm, angle=-90]{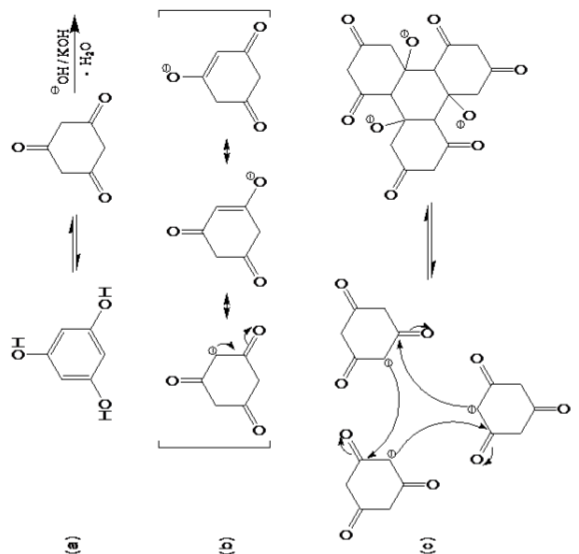}
  \caption{ Proposed synthesis of GO (a.) 1,3,5 - trihydroxybenzene a phenol-like substance whose oxo- and enol- forms  represent our monomer; (b.) deprotonation in typical for aldol-condensation conditions; (c.)  polycondensation in a hexagonal planar lattice.  \label{fig2}}
\end{figure}

\section{Results and discussion}

The particulars behind the experimental test of the general framework are dependent on the choice of the functional groups A and B reacting to form the lattice pattern of graphene or GO. One possible choice for the monomer  in a proof-of-concept experiment is 1,3,5 - trihydroxybenzene (phloroglucinol) given in Fig. \ref{fig2}a and the chemical reactions behind the polymerization (more appropriate here is  polycondensation) are nucleophilic additions to carbonyl groups. 

\begin{figure}[h]
\centering
  \includegraphics[height=6.2cm]{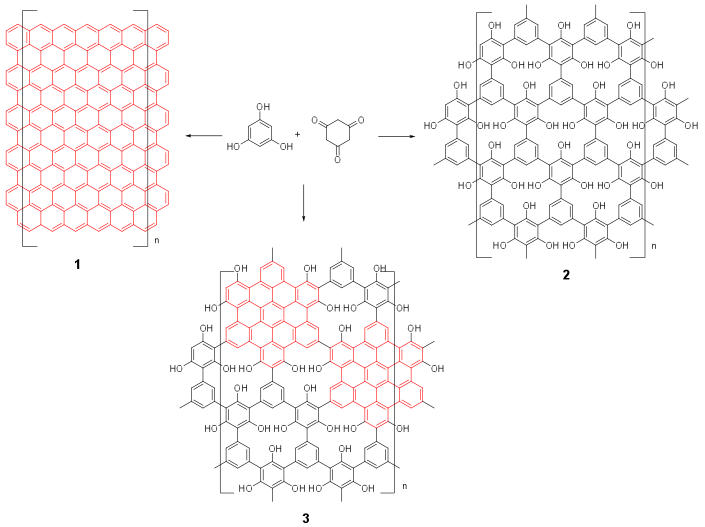}
  \caption{ Potential final structures of phloroglucinol two dimensional polymerization under typical aldol-condensation conditions. \label{structure}}
\end{figure}

The feasibility of the two dimensional polymerization using this monomer and GO in mind as the final product must be addressed in terms of the requirements discussed in \citep{poly1}. The solution based approach here does not depend on the ordering power of an interface, therefore the symmetry group $C_{3\nu}$ of the phloroglucinol ensures growth into the hexagonal carbon lattice of the same $C_{3\nu}$ group in the foundation of graphene and GO. Phloroglucinol has 3 functional groups and 3 latent sites in alternating positions capable of bond formation to the functional groups of other molecules.  Coupling reactions can occur in the same plane since the monomers' functional groups in play during the bond-formation event are sensitive to whether or not the fragments are coplanar (see Fig. \ref{fig2}c). Holes are formed when to fragments with incompatible edges join. These events do not destroy the two dimensional character of the resulting polymer. In principle, the most important critical factor to the solution approach is the enormous loss of solubility with the increasing size of the two dimensional fragments.  In our case, this problem is overcome by the use of a monomer with low solubility  to end up with a polymer with sufficient solubility, that is GO (unlike graphene).

Phloroglucinol has a unique symmetrical structure corresponding to the two tautomeric forms. This molecule is a potential building block for two dimensional polymers, because of its alternating three nucleophilic and three electrophilic centers \citep{pol2} and can give two-dimensional products under aldol-condensation condition. The two dimensional aldol condensation between oxo- and enol- form can have two ideal directions - complete condensation leading to a graphene {\bf 1} or two-dimensional polymer {\bf 2 \& 3} (2DP), see Fig. \ref{structure}.

\begin{table}
  \centering
  \caption{ \footnotesize{ The $2 \theta$ at CuK$\alpha$ peak list of the synthetic GO2 compared to the peak lists of commercial rGO by NanoInnova\citep{NanoInnova} and Ruoff's group GO\citep{Murali}. A deviation of $.5$ deg is marked ''close'', a deviation of less than $.5$ deg is marked ''ok'' and no deviation is marked ''perfect.'' The absence of the GO peak in the range of 10-12 deg assigned to (002) plane confirms the true two dimensional nature of the synthetic GO. Standard procedures of producing GO are unable to fully exfoliate graphite and few layer crystallites are always present thus peaks from (00n) planes are strong. } }
  \label{xrd2}
  \begin{tabular}{llll}
    \hline
   \footnotesize{synthetic GO}	& \footnotesize{ rGO NanoInnova} &	\footnotesize{Ruoff's group GO} & \footnotesize{Match} \\
    \hline
\footnotesize{ n.a. } & \footnotesize{ n.a. } & \footnotesize{ 10 } & \footnotesize{ n.a. }\\
 \footnotesize{16.2 $\pm$ 0.2 }& \footnotesize{n.a. }& \footnotesize{15.5 }& \footnotesize{close} \\
 \footnotesize{20.4 $\pm$ 0.2 }& \footnotesize{20.4 }& \footnotesize{20.3 }& \footnotesize{perfect} \\
 \footnotesize{22.4 $\pm$ 0.3 }& \footnotesize{23.7 }& \footnotesize{n.a. }& \footnotesize{ok} \\
 \footnotesize{25.9 $\pm$ 0.2 }& \footnotesize{25.9 }& \footnotesize{n.a. }& \footnotesize{perfect} \\
 \footnotesize{28.6 $\pm$ 0.1 }& \footnotesize{n.a. }& \footnotesize{28.9 }& \footnotesize{ok} \\
 \footnotesize{30.2 $\pm$ 0.1 }& \footnotesize{30.3 }& \footnotesize{30 }& \footnotesize{perfect} \\
 \footnotesize{33.0 $\pm$ 0.2 }& \footnotesize{32.1 }& \footnotesize{33.1 }& \footnotesize{ok} \\
 \footnotesize{38.7 $\pm$ 0.4 }& \footnotesize{n.a. }& \footnotesize{n.a. }& \footnotesize{n.a.} \\
 \footnotesize{40.8 $\pm$ 0.1 }& \footnotesize{n.a. }& \footnotesize{40.4 }& \footnotesize{close} \\
 \footnotesize{42.5 $\pm$ 0.3} & \footnotesize{n.a. }& \footnotesize{42.5 }& \footnotesize{perfect} \\
 \footnotesize{43.4 $\pm$ 0.2}  & \footnotesize{43.4 }& \footnotesize{n.a. }& \footnotesize{perfect}\\
    \hline
  \end{tabular}
\end{table}

\begin{figure}[h]
\centering
  \includegraphics[width=6cm]{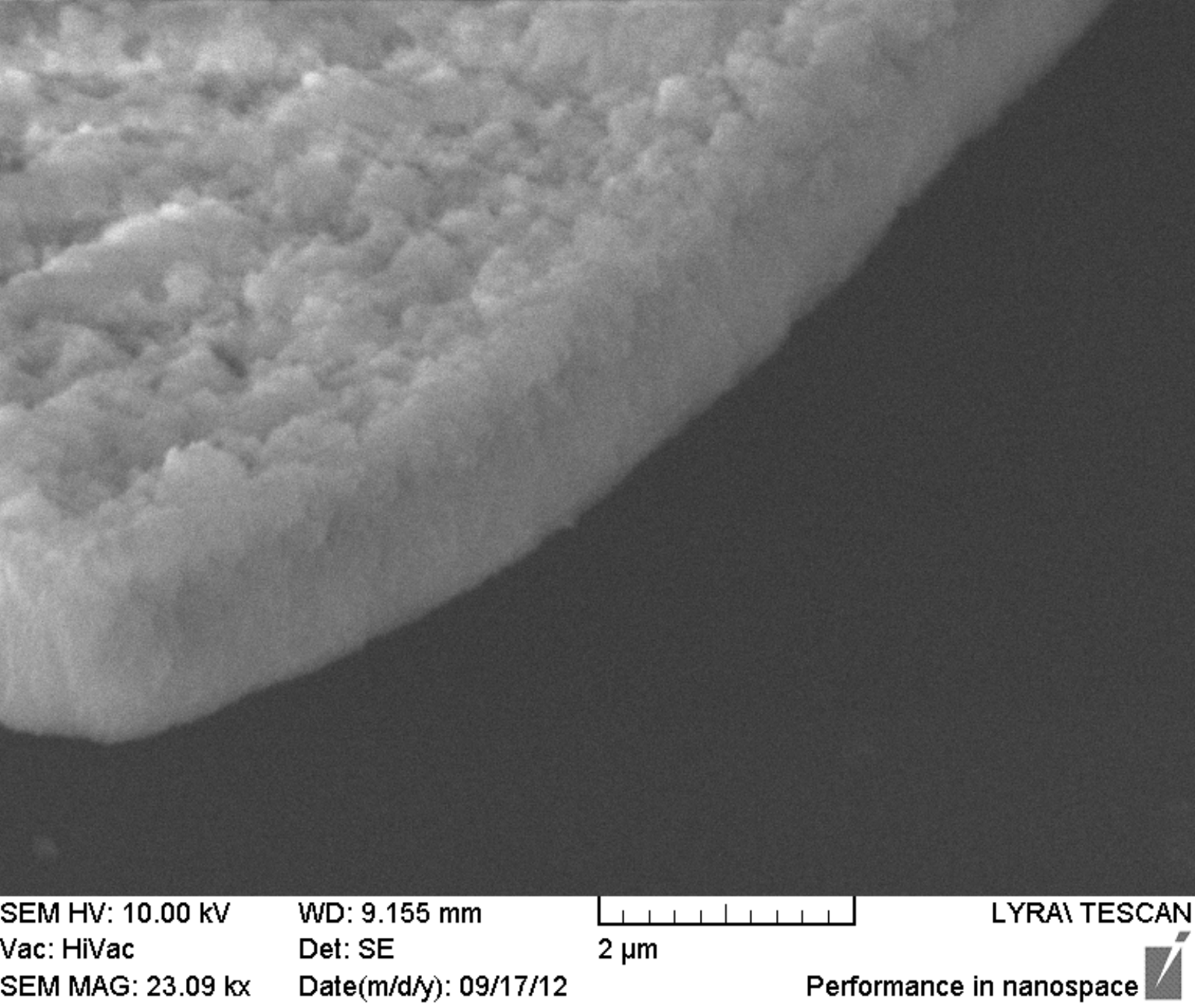}
  \caption{SEM image of the stacked up synthetic GO1 flakes $\sim 1$ $\mu$m in diameter. These formations, held up by hydrogen bonds originating from the hydroxyl groups attached to the hexagonal carbon planes, precipitate. \label{sem1}}
\end{figure}

\begin{figure}[h]
\centering
  \includegraphics[width=6cm]{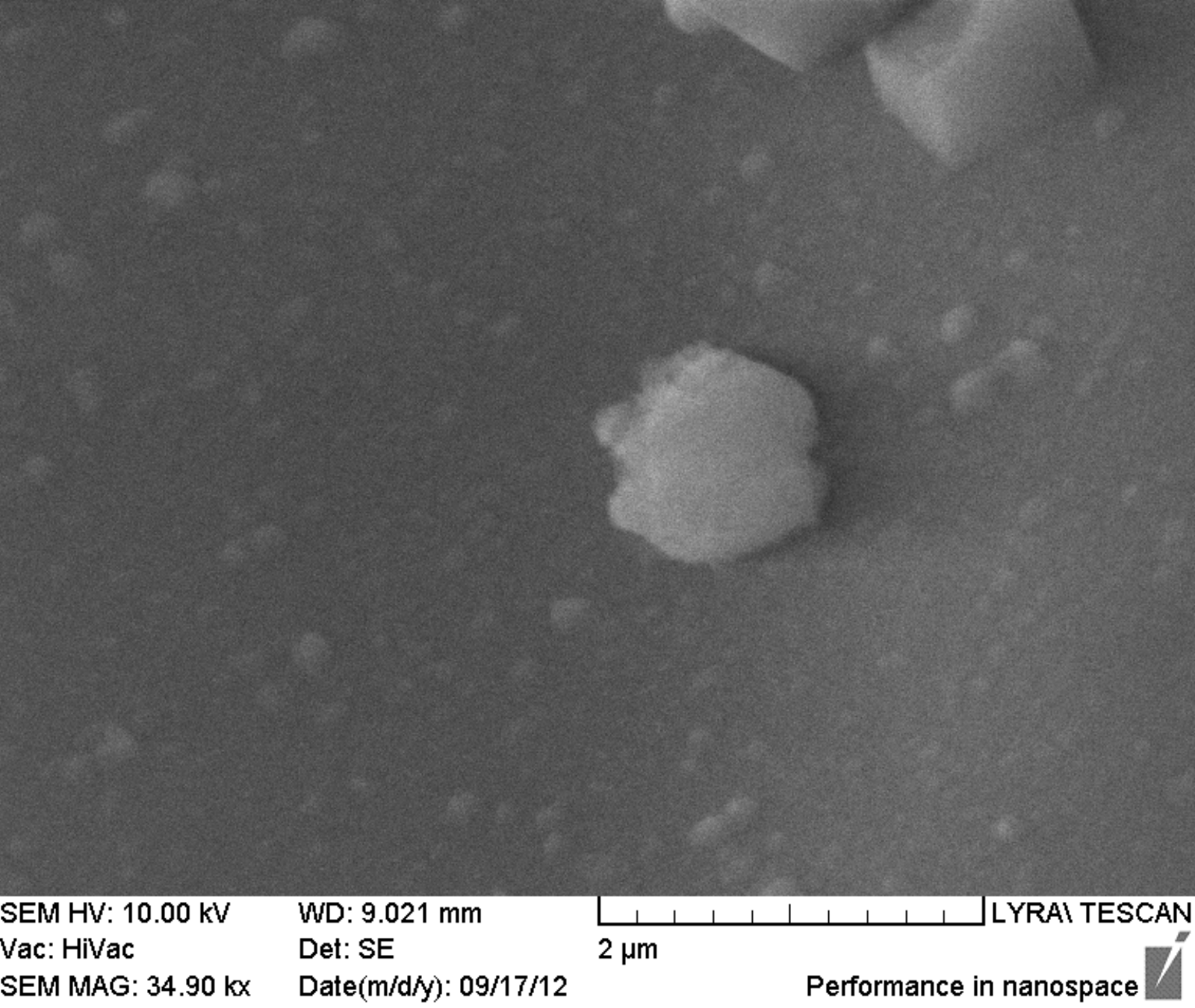}
  \caption{SEM image of a smaller (GO1) flake in the process of the stacking up. The hexagonal symmetry of the underlying lattice is noticeable. \label{sem2}}
\end{figure}

We tried similar approaches on typical for aldol-condensation conditions  (see Fig \ref{fig2}). The first approach was under mild conditions - room temperature with high amount of the base (potassium hydroxide) and low concentration of phloroglucinol (GO1). In this case we observed during the reaction time of four weeks small amounts of precipitate. We expected to obtain a product with the structure of  {\bf 1}, {\bf 2} or {\bf 3} (see Fig. \ref{structure}). The structure  {\bf 3} that we suggest (two dimensional polymer with graphene-like fragments) should be highly soluble in water and especially in basic solutions and the solubility will depend on the degree of condensation. The solubility is a function of the presence of hydroxyl groups which lead to the formation of secondary crystalline structure (hydrogen bonded) of the precipitate clearly visible in Fig. \ref{sem1} and Fig. \ref{sem2}.  The isolation yield  $<$ .05 \% of the reaction is possible to be explained with the high solubility of the final product. The yield we convey is in terms of mass since we have no information regarding the degree of polymerization as the MALDI-TOF experiments were unsuccessful. Therefore, we believe the molar mass of the resulting two dimensional polymer is enormous as confirmed by TEM and AFM imaging. Similar results we obtained also in higher temperature (refluxing in water) for a 72 h with the catalytic amount of potassium hydroxide. The yield in this case slightly increases $<$ 1\% but the size of the crystallites diminishes.

\begin{figure}[t]
\centering
  \includegraphics[width=5cm]{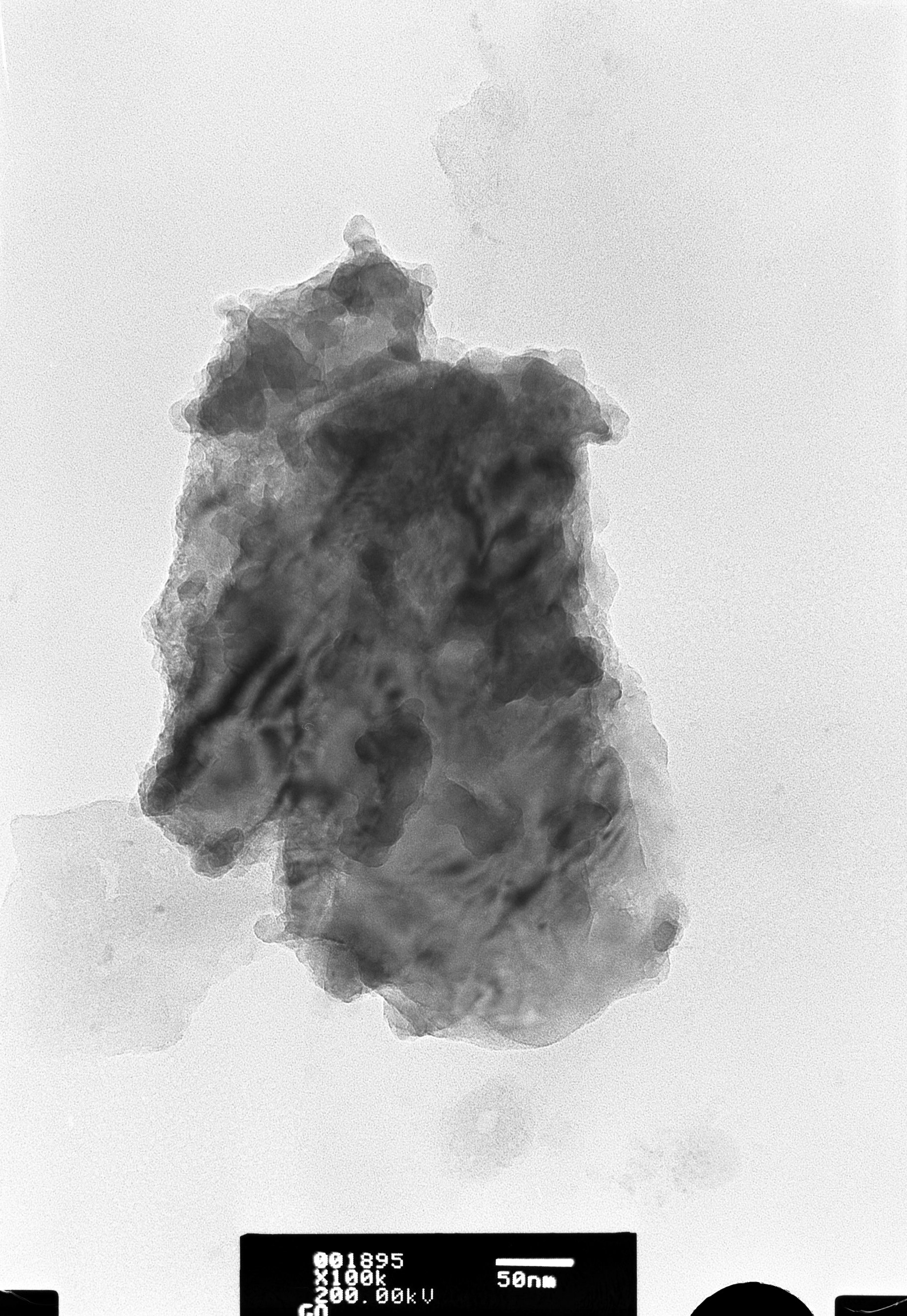}
  \caption{TEM image of synthetic GO1 flakes. The scale bar is 50 nm; The area of the flake is $\sim 1 \; \mu$m$^2$ rendering the flakes inaccessible for a MALDI-TOF mass spectrometry. \label{tem1}}
\end{figure}

The conceived chemical structures of the graphene oxides comprise those of Hofmann, Ruess, Scholz-Boehm, Nakajima-Matsuo  and Lerf-Klinowski \citep{gr3}. Therefore, there is  no unambiguous structure of the graphene oxide. In this paper we have assumed that if a basic list of criteria is fulfilled then the material could be identified as graphene oxide. These criteria are i.) 2d crystalline structure (provable with TEM, SAED, Raman, XRD) including "silk-like" sheet corrugations (see Fig. \ref{tem1}); ii.) thickness (1 - 1.2 nm) provable with AFM; iii.) oxygen content  and kind of C-O bonding (provable with XPS). We have also expanded the list of structures for the sake of the interpretation of the outcome of the bottom-up synthesis presented on Fig. \ref{structure}.

The structure and morphology of graphene oxide-like crystallites were explored by Raman spectroscopy, AFM, XPS, SEM, TEM, selected area electron diffraction pattern (SAED) and cathodoluminescence (CL). Crystalline structure was also characterized by X-ray diffraction (XRD). The potential structure {\bf 3} (see Fig. \ref{structure}) can be confirmed by the finding of islands of  sp$^2$ hybridised carbon (graphene like) and islands of hydroxyl groups arranged in groves on the synthetic flakes. Both structures are found on the synthetic GO1\&2 (see Experimental section) flakes as discussed below. Experimental data in the paper is provided for GO1.
\begin{figure}[h]
\centering
  \includegraphics[height=7.2cm]{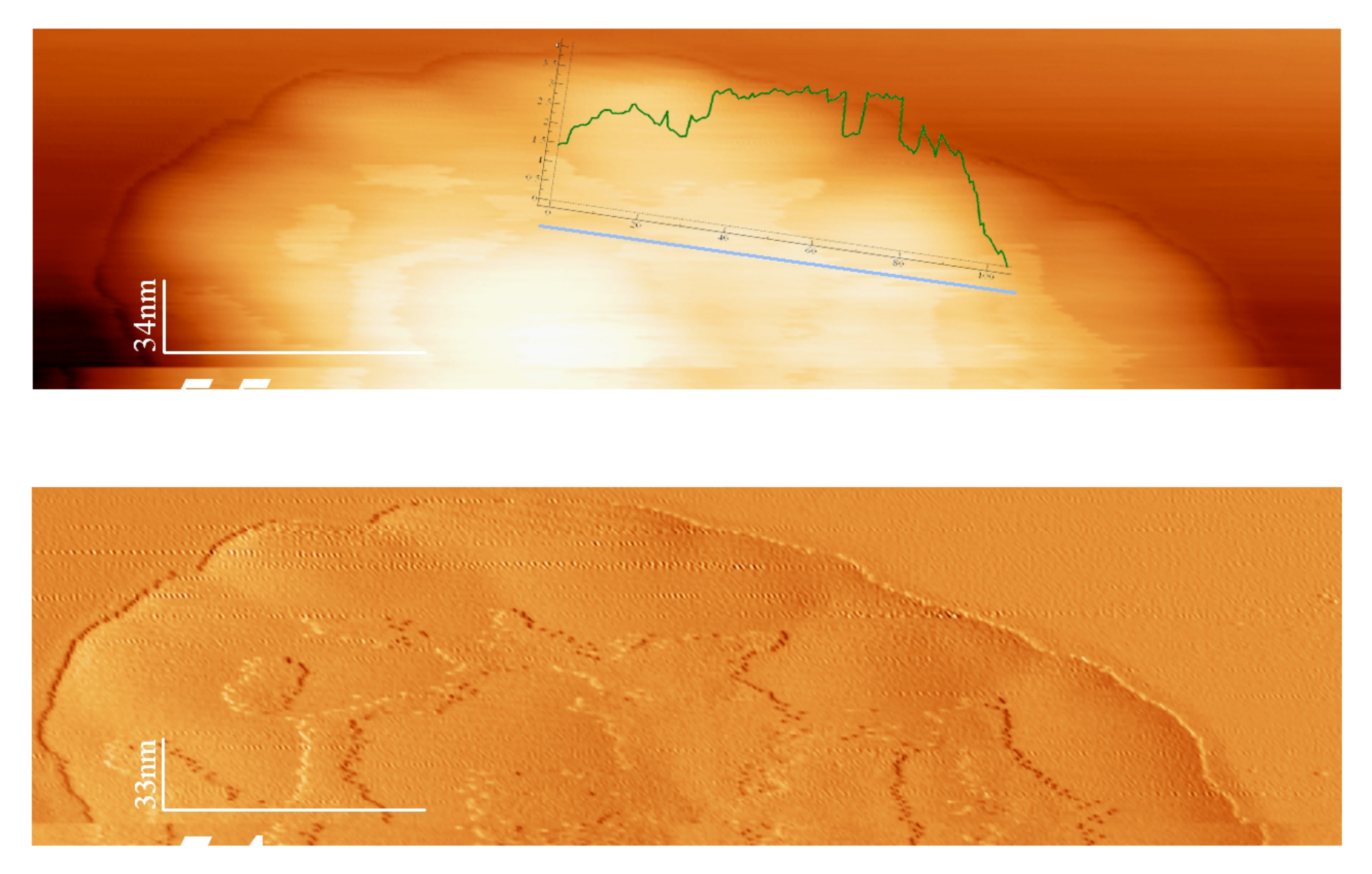}
  \caption{The AFM image of the graphene oxide flakes produced according to recipe GO1. The layer formations are $\sim 1$nm thick as expected for graphene oxide. Note the size of the crystallites exceeds $700 \; {\rm nm}^2 \approx 10^{-3} {\rm \mu m}^2 .$ Such crystallites contain over $10^5$ carbon and oxygen atoms rendering their molecular mass inaccessible for MALDI-TOF mass spectrometry. \label{fig8}}
\end{figure}

\begin{figure}[h]
\centering
  \includegraphics[width=8cm]{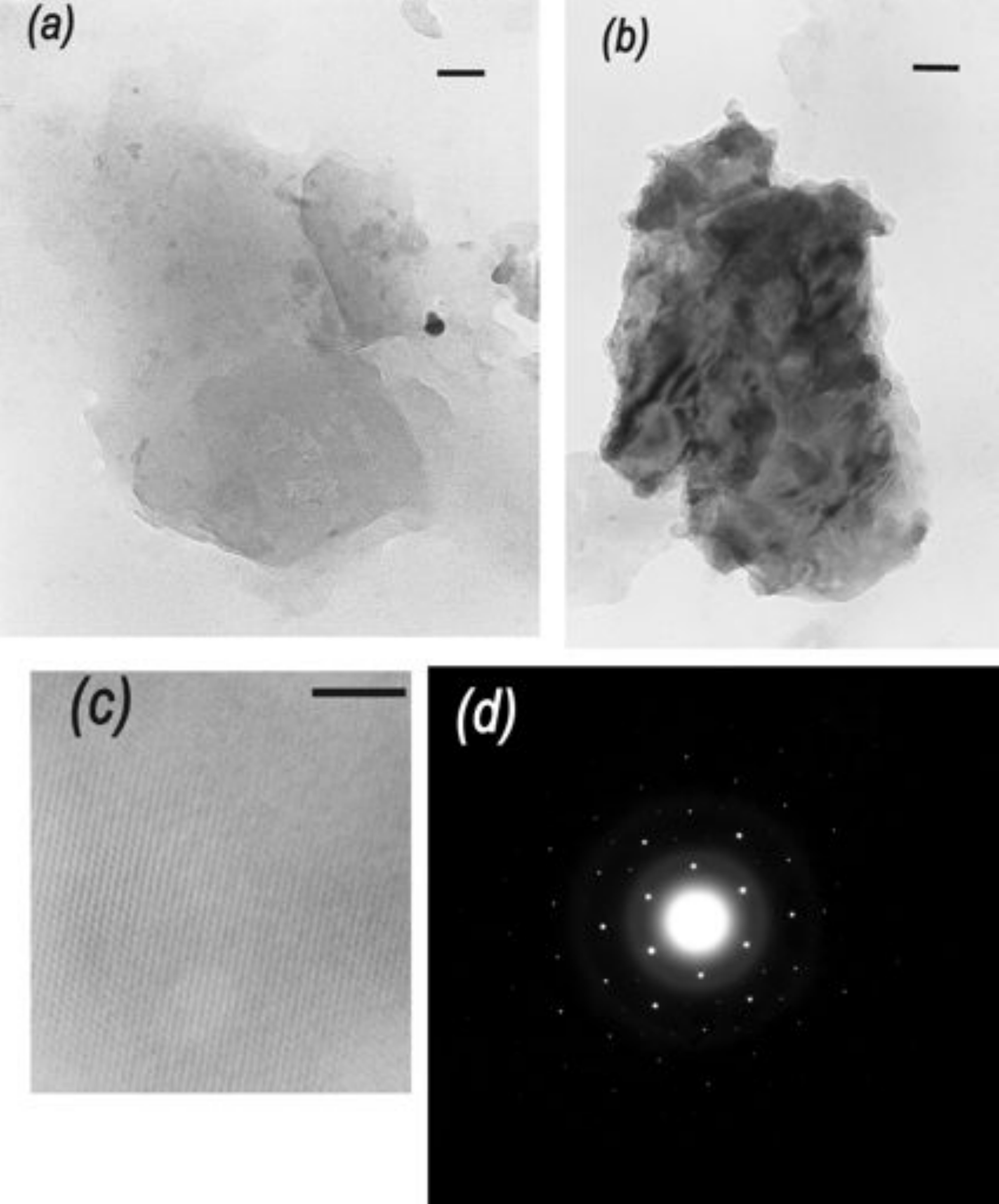}
  \caption{TEM images of synthetic GO1 flakes. The scale bar in (a.) \& (b.) is 50 nm; Scale bar in (c.) is 5 nm; (d.) The SAED confirms the carbon hexagonal lattice of graphene islands in GO and the two dimensional space group $p31m$. The spacings are d$_{10}$=(2.50 $\pm$ 0.27 ) $\AA$ and d$_{11}$= (1.47 $\pm$ 0.13 ) $\AA.$  The theoretical ratio $\Delta={d_{10} }/{ d_{11} } $ for this lattice $\Delta=\sqrt{3}$ is confirmed by the experiment $\Delta_{exp}=1.70$ with $2\%$ accuracy. \label{fig4}}
\end{figure}

\begin{figure}[h]
\centering
  \includegraphics[height=5.6cm]{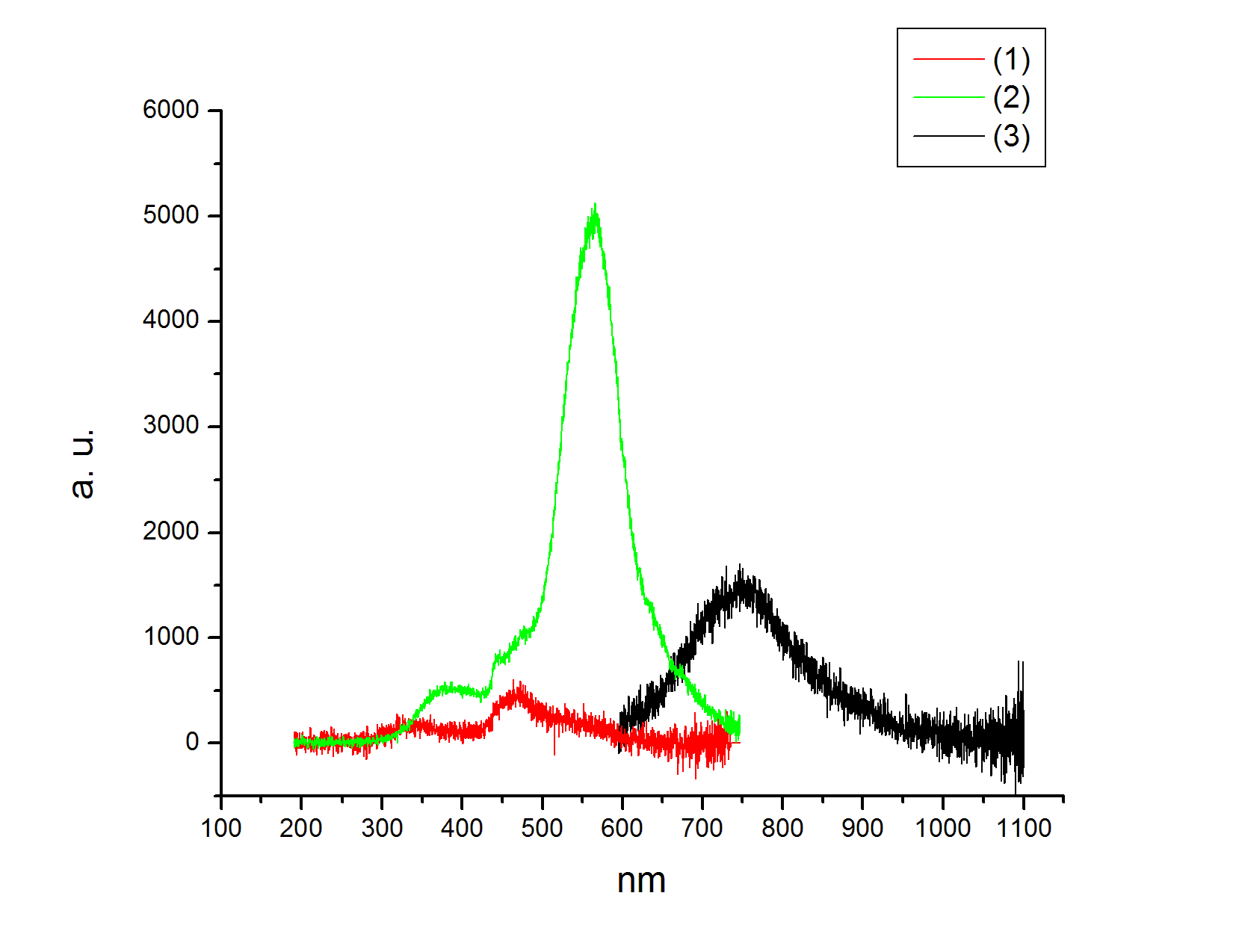}
  \caption{The cathodoluminescence spectra. The electrons in the SEM are accelerated toward the anode under potential differences of 25kV, and their current is 180 $\mu$A. The spectroscopic curves correspond to (1) Phloroglucinol; (2) Graphene Oxide - synthesized at room temp (GO1). ; (3) Graphene Oxide - refluxing in water (GO2). The band gap of the corresponding synthetic graphene oxides is (1) $E_{gap} \approx 1.9 \; {\rm eV} $ and $E_{gap} \approx 1.5 \; {\rm eV} .$ This lowering of the band gap in the second sample is a result of the increase of the C/O ration as confirmed by the XPS study. The extreme case is graphene where the degree of oxidation is vanishing as well as the band gap. \label{CL}}
\end{figure}

The AFM and TEM imagery as well as SAED for the first recipe GO1 (see Experimental section) are conveyed in Fig. \ref{fig8} and Fig. \ref{fig4}. The cathodoluminescence spectra is visible on Fig. \ref{CL}. The $\sim 1$nm thickness of the synthetic GO1 platelets is uniform as visible on the AFM image. This is in accordance with previous experimental results on graphene oxide \citep{AFM1, AFM2}. The two dimensional character of the formations where the two dimensional polymerization takes place is also visible. The estimated size of the GO nano-platelets is $700$ nm$^2$ which results in large potential sp$^2$ hybridized graphene-like islands. The SAED pattern was used to estimate the lattice spacings of the synthetic GO1. The results agree with the lattice spacing of graphene pointing to the sp$^2$ domains in the structure of the synthetic GO producing the diffraction pattern (the interplanar spacings are d$_{10}$=(2.50 $\pm$ 0.27 ) $\AA$ and d$_{11}$= (1.47 $\pm$ 0.13 ) $\AA$  where the theoretical ratio $\Delta={d_{10} }/{ d_{11} }= \sqrt{ 3} $ for this lattice belonging to the hexagonal space group $p31m$ is confirmed by the experiment $\Delta_{exp}=1.70$. The lattice parameters are $a=b=d_{10},$ $c=\infty,$ $\alpha=\beta=\pi/2$, $\gamma=2\pi/3$.). 

\begin{figure}[h]
\centering
  \includegraphics[width=8cm]{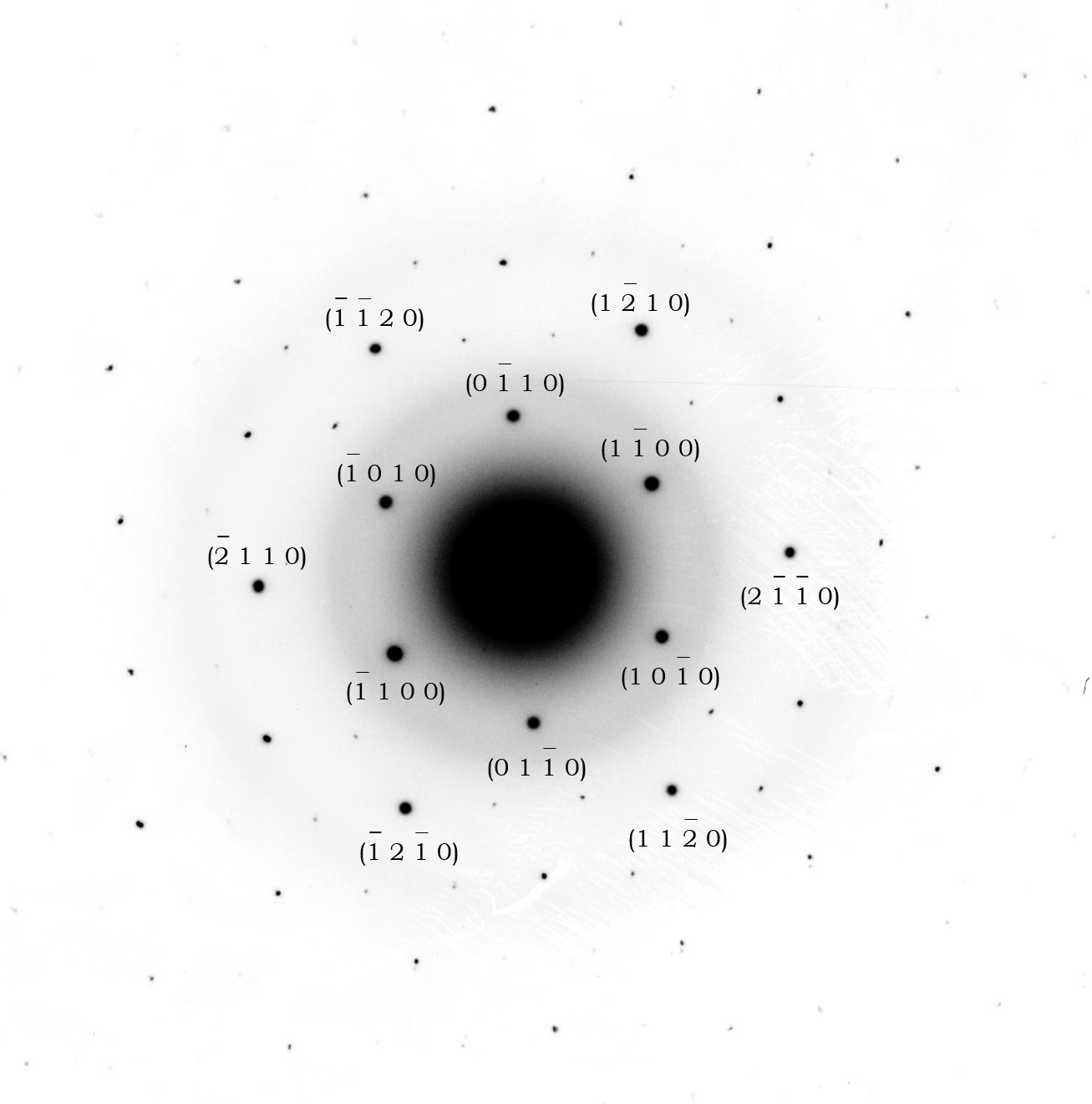}
  \caption{ The indexed SAED pattern (GO1 flake) confirming the two dimensional hexagonal carbon lattice belonging to the space group $p31m.$ \label{tem4}}
\end{figure}

\begin{figure}[b]
\centering
  \includegraphics[width=5cm]{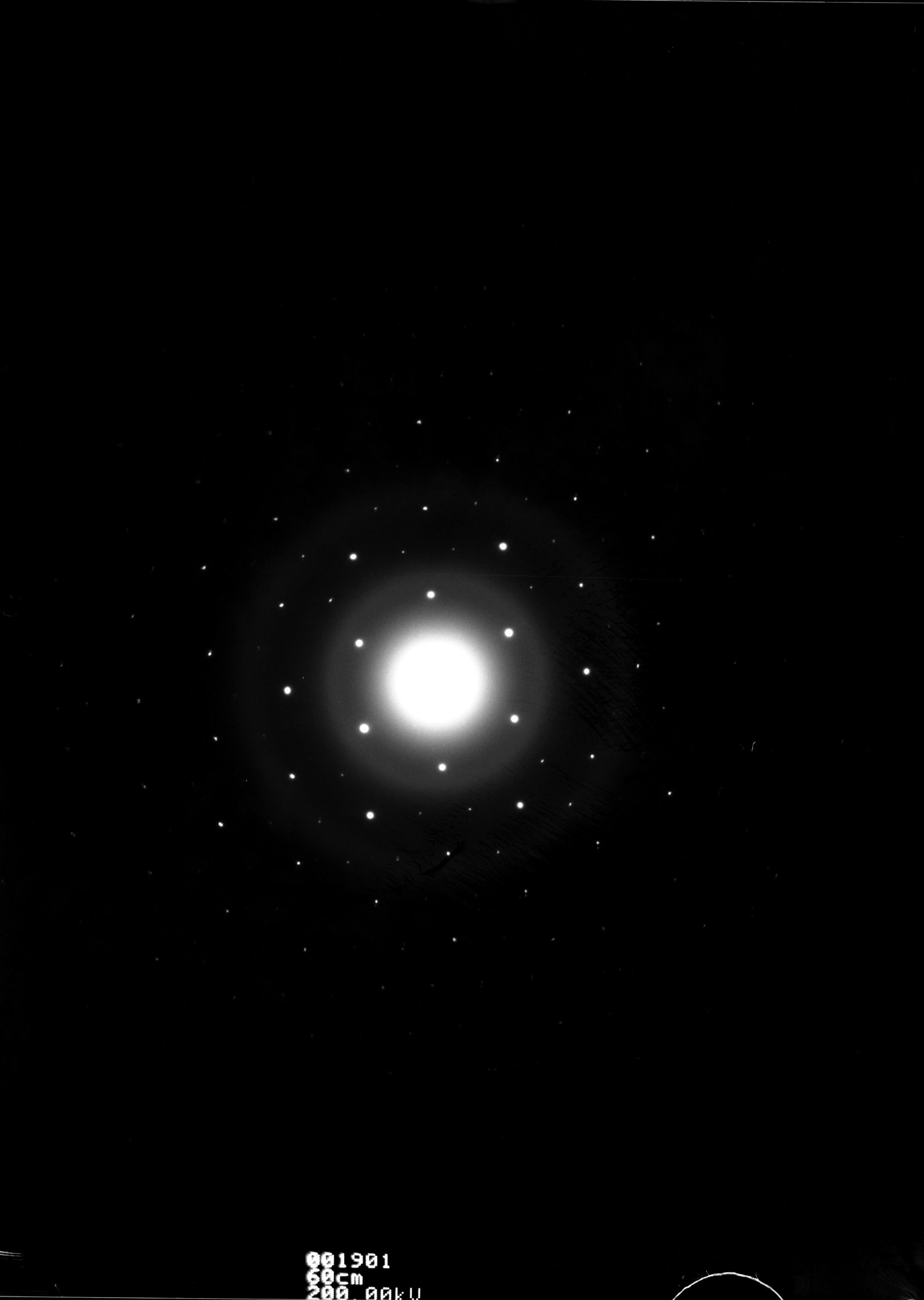}
  \caption{TEM image of synthetic GO1 flakes.  The electron diffraction image of the flakes confirms the carbon hexagonal lattice of graphene. The interplanar spacings are d$_{10}$=(2.50 $\pm$ 0.27 ) $\AA$ and d$_{11}$= (1.47 $\pm$ 0.13 ) $\AA.$  The theoretical ratio $\Delta={d_{10} }/{ d_{11} }= \sqrt{ 3} $ for this lattice is confirmed by the experiment $\Delta_{exp}=1.70$ \label{tem3}}
\end{figure}

\begin{figure}[t]
\centering
  \includegraphics[width=5cm]{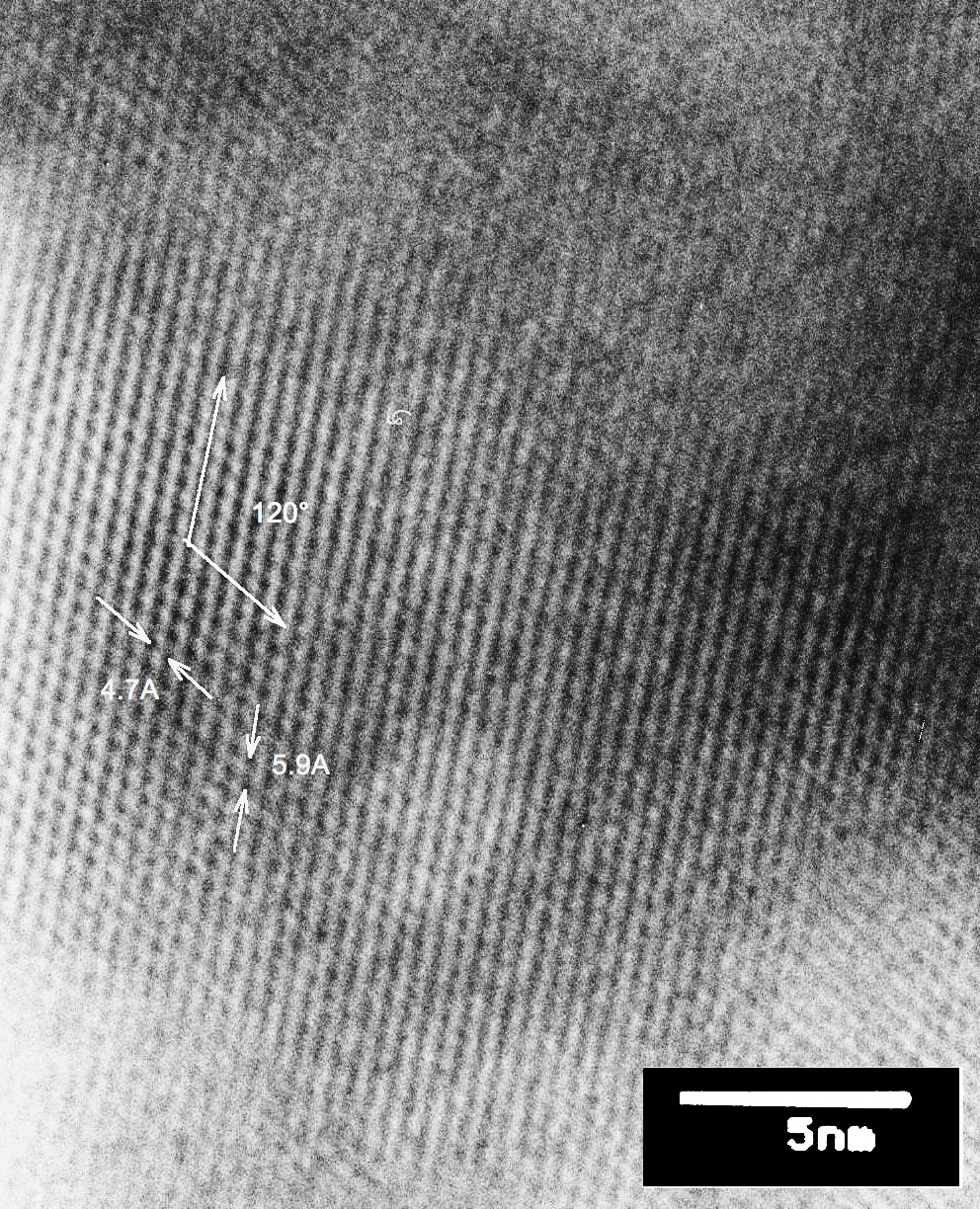}
  \caption{TEM image of synthetic GO1 flakes. The scale bar is 5 nm; The hydroxyl groups arranged in groves at $120$ deg. The lattice spacing in this secondary hydroxyl structure are $a=4.7 \AA$ and $b=5.9 \AA$. These spacings are not rigid and vary over $\sim 1 \AA$  in different positions over these sections probably as a function of the curvature of the sheet. \label{tem2}}
\end{figure}

Such a clear diffraction pattern can be produced from large domains of the order of at least few tens of nanometers in diameter, that is few hundreds, probably thousands of aromatic rings. The large sized sp$^2$ graphene-like domains (see Fig. \ref{tem4} \& \ref{tem3}) are also confirmed in the interpretation of the cathodoluminescence spectra. Interestingly, the TEM imagery reveals areas on the flakes larger than 100 nm$^2$ of ordered hydroxyl groups arranged in groves $\sim 4.7 \AA$ apart. The model for the possible bonding sites of --O and --OH on a graphene oxide layer\citep{ModelOH}, shows an arrangement where top and bottom --O and --OH groups are attached to the graphene sheet in a periodic fashion (see Fig. \ref{tem2}). Such a structure has been observed in STM images of GO prepared following the classic Hummers and Offemans method\citep{STMgo}. The lattice constants of this secondary crystalline structure of the synthetic GO1 extracted from the TEM image Fig. \ref{fig4}c are $a \sim 4.7 \AA$ and $b \sim 5.9 \AA$ which is a good agreement with previous findings\citep{ModelOH, STMgo}. The cathodoluminescence spectra is used to determine a property of the material, that is the band gap. In carbon materials containing a mixture of sp$^2$ and sp$^3$ hybridized bonds such as GO the opto-electronic properties are determined by the $\pi$ states of the sp$^2$ sites\citep{rob}. The luminescence of such carbon systems is a result of the recombination of localized e-h pairs on the sp$^2$ domains which behave as luminescence centers or chromophores\citep{heitz}. Since the band gap depends on the fraction, size and shape of the sp$^2$ domains, it varies in the synthetic GO. Based on a study of the band gap as a function of the sp$^2$ domain size\citep{goki} , we argue that these domains in the synthetic GO are larger than 37 aromatic rings, that is larger than previously synthesized graphenes\citep{mullen, mullen2}. This argument is reinforced by the clear SAED pattern of graphene-like domains of large area. 

The XRD peak list of the synthetic GO produced according to the second recipe GO2 (see Experimental section) in Table \ref{xrd2} is compared to the peak lists of chemically exfoliated GO of other research groups. The observed good agreement conveys the similar structure.

The expected Raman spectra signatures are the D, G, 2D and D + D' peaks \citep{g_raman}. In our spectra of synthetic GO1  G peak is at 1600 cm$^{-1}$ and D peak appears at 1370 cm$^{-1},$ see Fig. \ref{raman}.  The second order of the D peak, 2D (or G') peak position are also present in the range 2640- 2700 cm$^{-1}$ depending on the number of layers and the D + D' peak at 2940 cm$^{-1}$ is due to the defect activated combination of phonons which confirms the defective structure produced during the two dimensional polymerization.  Additionally, a peak at 1450cm$^{-1}$ is present and ts origin is attributed to a C-OH mode (phenol -OH group) and the characteristic medium band of the carbon ring.

In order to get an insight into the concentration of the functional groups, we performed a curve fitting procedure of carbon C1s peak as well as O1s peak in the X-ray Photoelectron spectra (Fig. \ref{xps}). The peaks of C1s spectra are assigned to four components that correspond to carbon atoms in different functional groups: the C in nonoxygenated ring (C-C), the C in C-O bonds (C-OH), the carbonyl C (C=O), and the carboxylate carbon (O=C-OH). The XPS of C 1s and O 1s  were obtained for the precursor and the synthetic GO via two methods of preparation GO1\&2. The C/O ratios are: phloroglucinol $C_{1s}/O_{1s}=2$; GO1 - $C_{1s}/O_{1s}=1.7$ and GO2 - $C_{1s}/O_{1s}=3.45$ .  A consumption of the C=O bonds during the reaction is also observed. The ratios $\alpha$ between the $C-O/C=O$ bonds as measured via the area of the fitting curves are: phloroglucinol $\alpha_{phl}=0.52$; GO1 - $\alpha_{GO1}=3.34$ and GO2 - $\alpha_{GO2}=3.30$. This confirms the aldol condensation proposed as the mechanism behind the two dimensional polymerization of 1,3,5-trihydroxybenzene to GO. The higher carbon content in the second GO sample is an indication of higher degree of graphenization confirmed by the shift to lower band gap in the CL spectra (Fig. \ref{CL}). 

Next we focus the discussion on the content of the figures in the remainder of the paper. We have conveyed two SEM images (Fig. \ref{sem1} and Fig. \ref{sem2}) of the secondary crystallite formation GO1 sample exhibited. Since the secondary crystalline structure carries remnants of the symmetry of the GO flake, we are conveying Fig. \ref{sem2} as an additional proof of the underlying hexagonal order. The thickness of the  crust-like flakes on Fig. \ref{sem1} coincide with the average size $D$ of the GO flakes $D\approx 1 \mu m.$ We believe these are a result of GO flakes stacked up on top of each other along the flat side held by hydrogen bonds which stem from the hydroxyl (epoxy) groups attached to the GO surface as confirmed by the XPS analysis. This structure is yet another confirmation of the successful attachment of oxygen containing groups. A stronger evidence of their presence is the TEM image on Fig. \ref{tem2} where the spacing and arrangement of the grooves on top of the GO surface is exactly as to be expected if they were to be attached onto a carbon graphene-like lattice, namely the angle between the intersecting lines of groves is 120 deg and their spacing is approximately twice d$_{10}$ interplanar spacing and three times d$_{11}$ interplanar spacing. These spacing are established from Fig. \ref{tem3} containing the selected area electron diffraction image. The indexed SAED depicted on Fig. \ref{tem4} is conveyed in order to reinforce the conclusion that the underlying lattice is indeed hexagonal and made up of carbons where the C-C bond length coincides with 
d$_{11}$= (1.47 $\pm$ 0.13 ) $\AA$ interplanar spacing. Special attention is to be paid to Fig. \ref{tem1}. This general outlook of the synthetic GO flakes exhibits silk like folding which is an indirect proof of the two dimensional character of the synthetic flakes. If they were to be held together by interlayer bonding (spanning the third dimension) the structure would stiffen and no such folding would be observed. The two dimensional character is additionally confirmed by the X-ray diffraction patterns conveyed on Fig. \ref{xrd1_fig} and Fig. \ref{xrd2_fig}. Due to lack of sufficient amount of GO material we have mixed the powder with a known halide whose diffraction peaks are not superimposed onto the ones expected from the GO. An attempt at automatic identification of the material is depicted on Fig. \ref{xrd1_fig}. Clearly the pattern is such that a manual confirmation of the  structure is necessary which is done by extracting the relevant peaks (see Table \ref{xrd1}) and comparing them with the ones from known XRD studies of GO (see Table \ref{xrd2}).  Most importantly, no trace of interlayer bonding is found thus confirming the two dimensional character of the synthetic material.
More experimental data and figures are available in the Experimental section.

In summary, we have proposed a two dimensional polymerization for the synthesis of graphene oxide. A proof-of-concept experiment was conducted and the experimental evidence for the feasibility of this two dimensional polymerization conveyed. 

%%%%%%%%

\nopagebreak

\section{Experimental}

The synthesis of graphene oxide from phloroglucinol  in an alkaline solution was done in two different solution conditions.

GO1: The initial alkaline solution prepared in an ultrasonic bath was 4.25 wt. \% KOH (potassium hydroxide bought from Valerus Ltd. ( http://www.valerus-bg.com ),  with 85\% KOH) water solution (DI  18 M$\Omega.$cm) into which 0,1 wt. \% of phloroglucinol dihydrate (1,3,5 trihydroxybenzene C$_6$H$_3$(OH)$_3$.2H$_2$O from Carlo Erba) was introduced. The initial solution has a bluish purple tint characteristic of metal-phenol ion complexes. The mixture was left for 4 weeks in a closed with parafilm glass eprouvette in a semi-darkened room. The color of the mixture turns to yellow and a white semi translucent sediment appears at the bottom of the vessel. The sediment was separated in a centrifuge (10 000 rmp) in a series of 10 fomentations with DI. The procedure was a consecutive fomentation, centrifuge and removal of the water column above the sediment. This removed the water soluble salts.  An aliquota was taken on a silicon wafer for XPS and Raman spectroscopy,  An aliquota was taken on a TEM grid for TEM and SAED as well as on mica for AFM. An aliquota was taken on a metallic surface for cathodoluminescence. Yield $<$ .05\%

GO2: The second alkaline solution tested was refluxed at boiling temperature.  A mixture of phloroglucinol dihydrate (10 mmol, 1.62 g), potassium hydroxide (3.33 mmol, 0.22 g) was dissolved in water (50 ml). The mixture was refluxed for 72 h. After the reaction mixture cooled down it was acidified with hydrochloric acid (pH~0.5). The mixture was kept 7 days at room temperature and brown precipitate formed. The suspension was filtered through the glass filter G5 (porosity 1.0 --1.6 $\mu$m). The precipitate was washed several times with distilled water and dried. Yield $<$ 1\%.
An aliquota was taken for the above mentioned tests as well as for XRD which required greater amount of substance.

\begin{figure}[b]
\centering
  \includegraphics[height=4cm]{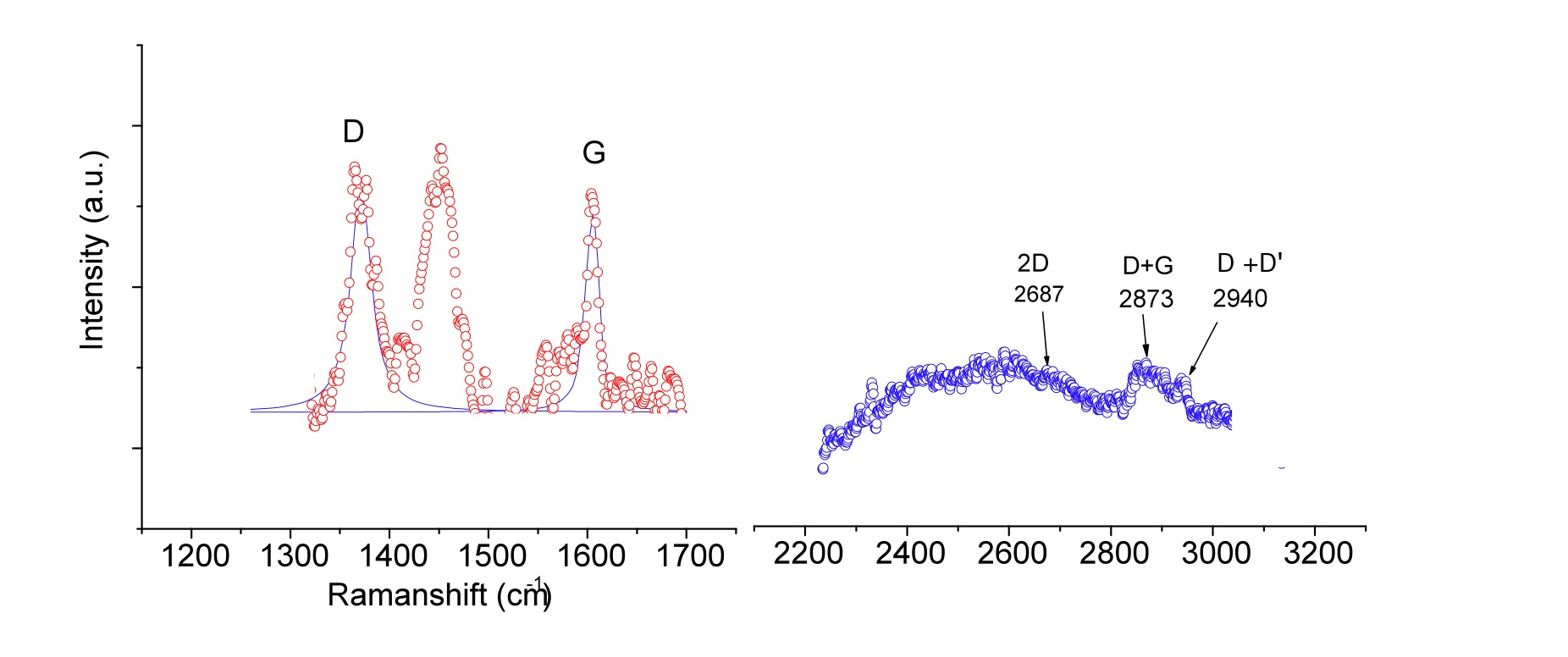}
  \caption{Raman spectra of synthetic GO1. \label{raman}}
\end{figure}

The scanning electron images were taken using a LYRA I XMU scanning electron microscope (Tescan).  The TEM observations and selected area electron diffraction (SAED) analyses were performed on a JEM2100 high resolution transmission electron microscope (HRTEM JEOL) operated at 200 kV.  The images were captured on a photographic plate 60x90mm.  The Raman measurements were carried out using micro-Raman spectrometer LabRAM HR800 Visible with He-Ne (633 nm) laser. At room temperature an objective $\times$100 was used both to focus the incident laser beam and to collect the scattered light.

\begin{figure}[b]
\centering
  \includegraphics[height=10cm]{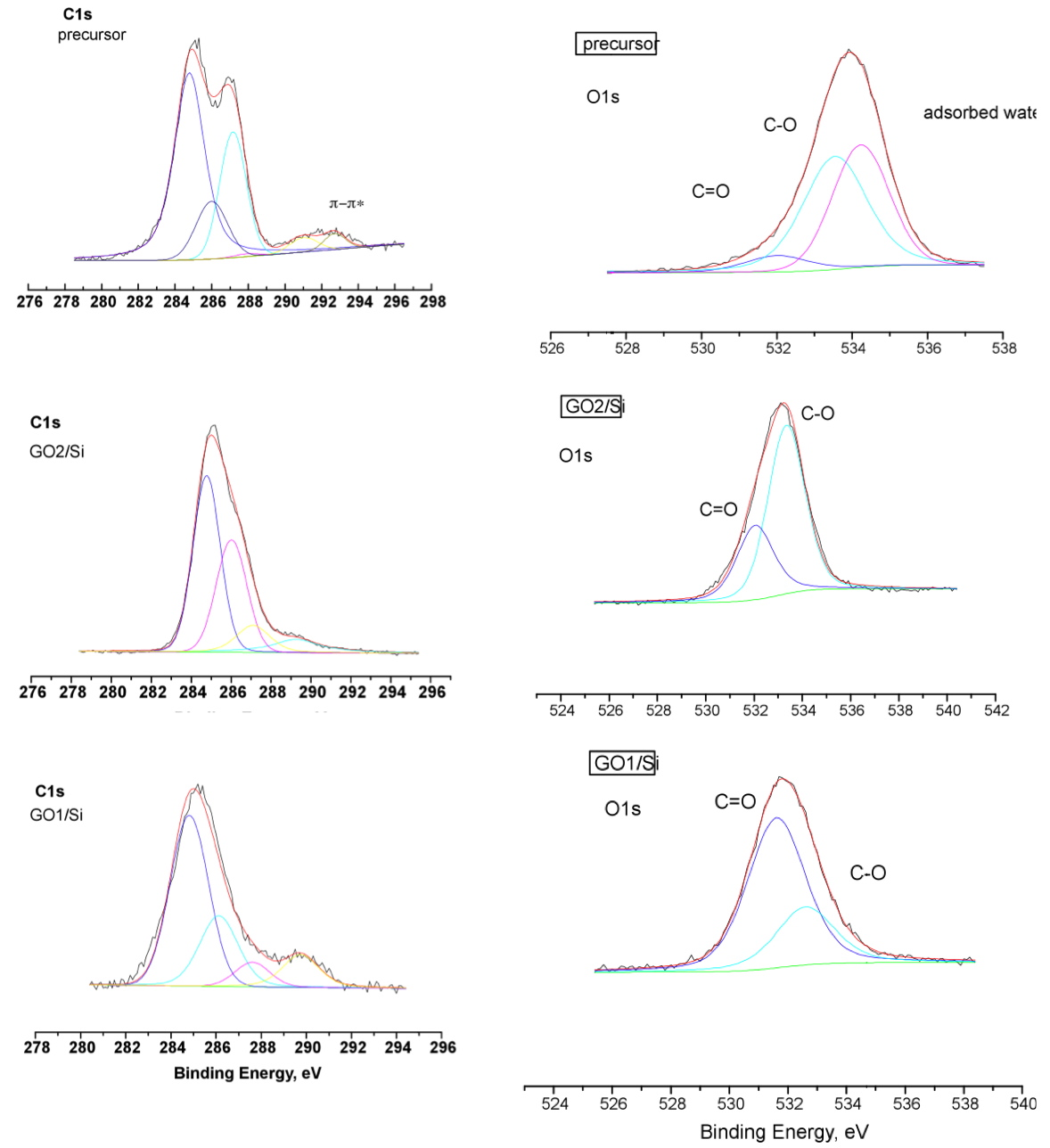}
  \caption{The XPS spectra of C1s and O1s in synthetic GO1\&2. \label{xps}}
\end{figure}

AFM imaging was performed on a NanoScopeV system (Veeco 
Instruments Inc.) operating in tapping mode in air at room temperature. We used silicon cantilevers (Tap300Al-G, BudgetSensors, 
Innovative solutions Ltd., Bulgaria) with 30-nm-thick aluminum 
reflex coating. According to the producers datasheet, the cantilever spring constant was in the range of 1.5 -- 15 N/m and the resonance frequency was 150 $\pm$ 75 kHz. The tip radius was less than 
10 nm.  The scan rate was set at 1 Hz, and the images were captured 
in the height and phase mode. Subsequently, all the images were flattened
by means of the Nanoscope software. 

The cathodoluminescence spectra were taken with an Avantes Spectrometer, Avaspec-2048 TEC-2 collecting light from the SEM Hitachi S-570 (3--30kV; up to 300$\mu$A electron current and 3.5nm ultimate spacial resolution) chamber through an optical fiber in a vacuum feedthrough.

\begin{figure}[h]
\centering
  \includegraphics[height=5cm]{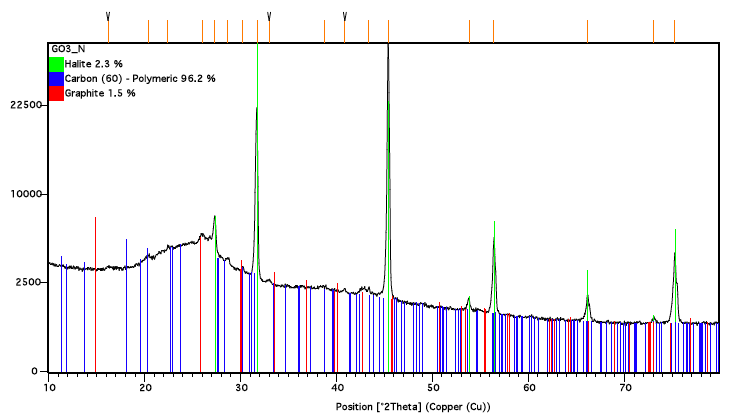}
  \caption{ The X-ray diffraction pattern with  Cu K$\alpha$ (1.54 A) of the synthetic GO2 mixed with NaCl. The pattern list of the matches includes NaCl (ref. code: 01-075-0306, the peak list for $2 \theta$ marked in green color includes 27.3 ; 31.7 ; 45.4 ; 53.8 ; 56.4; 66.1 ; 73.1 ; 75.3 degrees) which is indeed present. The remaining peaks at $2\theta$ are best fitted with Polymeric Carbon (ref. code: 98-005-6668) given in blue color and Orthorombic Graphite (ref. code: 98-008-8812) depicted in red color. The presence of the latter form of carbon in sp$^3$ hybridization is a strong argument in favor of graphene oxide in addition to the former match - the polymeric carbon.\label{xrd1_fig}}
\end{figure}

\begin{figure}[t]
\centering
  \includegraphics[height=5cm]{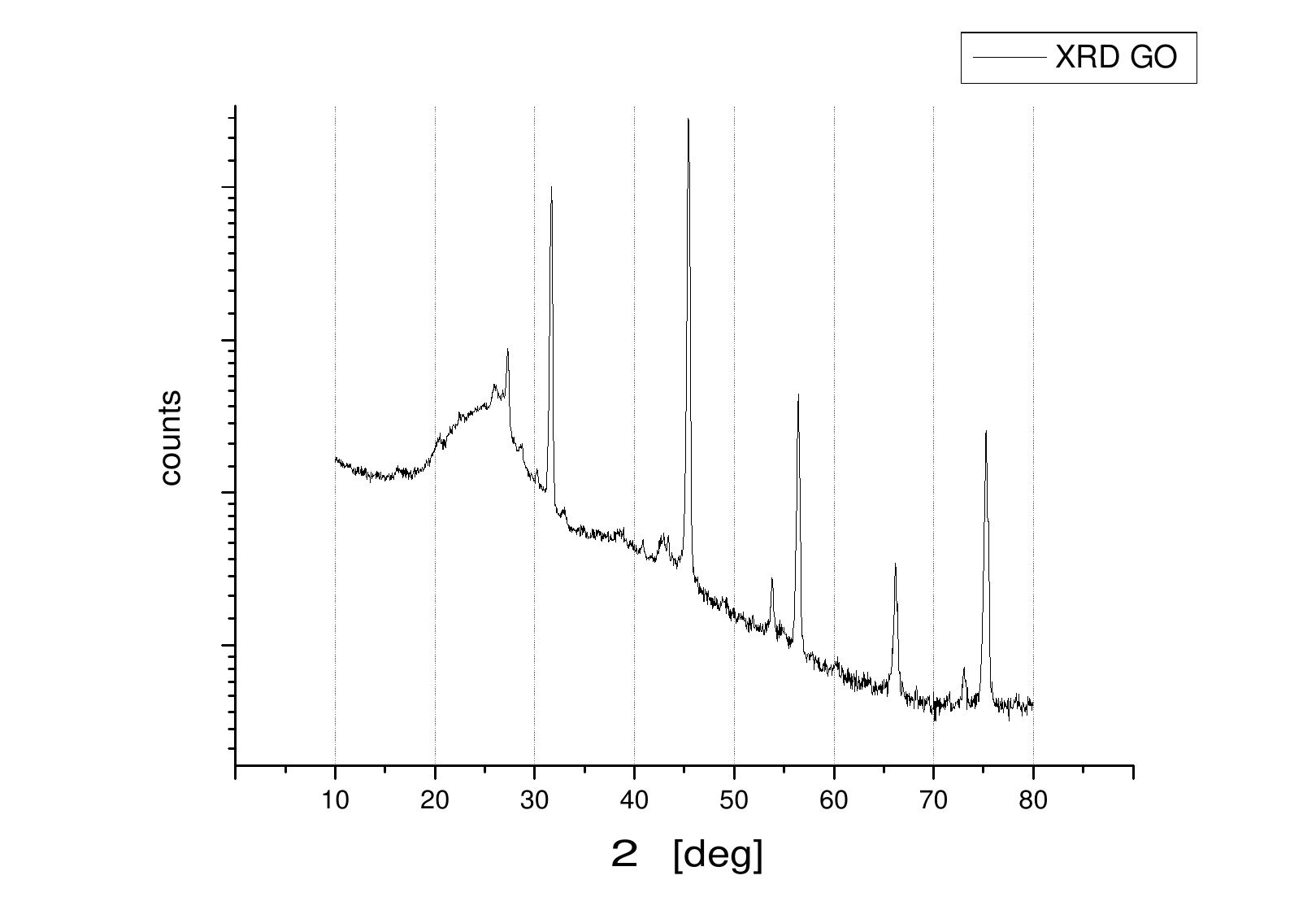}
  \caption{ The X-ray diffraction pattern with  Cu K$\alpha$ (1.54 $\AA$) of the synthetic GO2 mixed with NaCl. \label{xrd2_fig}}
\end{figure}

X-ray powder diffraction patterns for phase identification were recorded in the angle interval 10--80 deg (2$\theta$), on a Philips PW 1050 diffractometer, equipped with Cu K$\alpha$ tube, scintillation detector and monochromator in the diffracted beam. Data for cell refinements were collected in step-scan mode in the angle interval at steps of 0.03 deg and counting time of 10 s/step. The synthetic GO was mixed with a known substance NaCl to increase sample's volume and be able to take its diffraction pattern shown on Fig.\ref{xrd2_fig}. The identified characteristic $2\theta$ peaks for the synthetic GO are summarized in Table \ref{xrd1} and compared to $2\theta$ peaks for chemically exfoliated GO in Table \ref{xrd2}. The automatic pattern recognition using all known XRD databases is conveyed in Fig.\ref{xrd1_fig}.

The expected Raman spectra signatures measured at laser excitation $\lambda = 632.8$ nm are the D, G, 2D and D + D' peaks \citep{g_raman}. In our spectra Fig. \ref{raman} G peak shows hardening and reaches the highest position of 1600 cm$^{-1}$ and D peak appears at 1370 cm$^{-1}.$ The second order of the D peak, 2D (or G') peak position is usually observed in the range 2640- 2700 cm$^{-1}$ depending on the number of layers and the D + D' peak at 2940 cm$^{-1}$ is due to the defect activated combination of phonons. The wide spreading 2D band and the shoulder at higher wavenumbers indicate formation of multi-layer GO sheets. Additionally, a peak centered at 1450cm$^{-1}$ is modulating the spectrum. Its origin is attributed to a C-OH mode (phenol -OH group) and the characteristic medium band of the carbon ring \citep{spectro}.

%\begin{figure}[h]
%\centering
 % \includegraphics[height=7cm]{fig8_exp.pdf}
 % \caption{The AFM image of the graphene oxide flakes produced in the two dimensional polymerization. The layer formations are $\sim 1$nm thick as expected for graphene oxide. Note the size of the crystallites exceeds $700 \; {\rm nm}^2 \approx 10^{-3} {\rm \mu m}^2 .$ Such crystallites contain over $10^5$ carbon and oxygen atoms rendering their molecular mass inaccessible for MALDI-TOF mass spectrometry. \label{afm}}
%\end{figure}

%\begin{figure}[h]
%\centering
 % \includegraphics[height=7cm]{fig5_exp.png}
 % \caption{The cathodoluminescence spectra. The electrons in the SEM are accelerated toward the anode under potential differences of 25kV, and their current is 180 $\mu$A. The spectroscopic curves correspond to (1) Phloroglucinol; (2) Graphene Oxide - synthesized at room temperature; (3) Graphene Oxide - synthesized at solution's boiling point and a reflux condenser. The band gap of the corresponding synthetic graphene oxides is (1) $E_{gap} \approx 1.9 \; {\rm eV} $ and $E_{gap} \approx 1.5 \; {\rm eV} .$ This lowering of the band gap in the second sample is a result of the increase of the C/O ration as confirmed by the XPS study. The extreme case is graphene where the degree of oxidation is vanishing as well as the band gap. \label{CL}}
%\end{figure}

\begin{table}[t]
  \centering
  \caption{The $2 \theta$ [deg] at CuK$\alpha$ peak list of the synthetic GO2.}
  \label{xrd1}
  \begin{tabular}{llll}
    \hline
    Pos. $2 \theta$	& FWHM &	d-spacing [$\AA$]	& Rel. Int. [\%] \\
    \hline
16.2	& 0.5 & 5.48 & 0.53\\
20.4	& 0.5 & 4.36 & 0.82 \\
22.4	& 0.6 & 3.96 &	0.99 \\
25.9	& 0.5 & 3.43 &	4.18 \\
28.6	& 0.3 & 3.12 &	1.15 \\
30.2	& 0.2 & 2.96 &	0.89 \\
33.0	& 0.4 & 2.72 &	0.44 \\
38.7	& 0.8 & 2.33 &	0.43 \\
40.8	& 0.3 & 2.21 & 0.61 \\
42.5	& 0.6 & 2.13  & 1.40 \\
43.4  & 0.4 & 2.08 & 1.40 \\
    \hline
  \end{tabular}
\end{table}

The X-ray Photoelectron spectra were obtained using unmonochromatized MgK$\alpha$ (1486.6 eV) radiation in a VG ESCALAB MK II electron spectrometer under base pressure of 1x10-8 Pa. The spectrometer resolution was calculated from the Ag3d5/2 line with the analyzer transmission energy of 20 eV. The half-width of this line was 1 eV. The spectrometer was calibrated against the Au4f7/2 line (84.0 eV) and the sample charging was estimated from C1s (285 eV) spectra from natural hydrocarbon contaminations on the surface. The accuracy of the BE measured was 0.2 eV.
The photoelectron spectra of C 1s, O 1s were recorded and corrected by subtracting a Shirley-type background and quantified using the peak area and Scofield's photoionization cross-sections.

The concentration of the functional groups is inferred by a curve fitting procedure of carbon C1s peak as well as O1s peak using a Gaussian-Lorentzian peak shape (Fig. \ref{xps}). The peaks of C1s spectra are assigned to four components that correspond to carbon atoms in different functional groups: the nonoxygenated ring C(C-C), the C in C-O bonds (C-OH), the carbonyl C (C=O), and the carboxylate carbon (O=C-OH). The binding energy of the C-C and C-H bonding are assigned at 284.5-285 eV and chemical shifts of +1.5, +2.5 and +4.0 eV are typically assigned for the C-OH, C=O, and O=C-OH functional groups, respectively\citep{xps1, xps2}.  Also present in the first two investigated samples is a broad small peak between 290 and 292 eV, which corresponds to the $\pi$* shake-up transition associated with the aromatic ring.

Most structural models of GO also include an epoxide groups (C-O-C), which should have a C1s binding energy similar to C-OH \citep{xps3}. The peaks of O1s spectra are also assigned to C-OH, C=O and O=C-OH groups respectively. Shift to lower binding energy in O1s spectra is observed, depending on way of preparation of GO which indicates a transformation of C=O and C=O-OH as well as C-O groups to a new chemical species \citep{xps4}. We observe qualitatively and quantitatively that the concentration of carbon is increased in favor of oxygen in both recipes GO1\&2. 

The X-ray Photoelectron spectra of C 1s and O 1s (Fig. \ref{xps}) were obtained for the precursor  phloroglucinol, as well as for the synthetic GO via the two methods of preparation: GO1 - room temp. synthesis and GO2 - synthesis at boiling point and a reflux condenser. Both products were placed on top of a Si dice. The C/O ratios are as follows: Phloroglucinol $C_{1s}/O_{1s}=2$; GO1 - $C_{1s}/O_{1s}=1.7$ and GO2 - $C_{1s}/O_{1s}=3.45$ . A consumption of the C=O bonds during the reaction is also observed. The ratio $\alpha$ between the $C-O/C=O$ bonds as measured via the area of the fitting curves is: phloroglucinol $\alpha_{phl}=0.52$; GO1 - $\alpha_{GO1}=3.34$ and GO2 - $\alpha_{GO2}=3.30$. This confirms the aldol condensation proposed as the mechanism behind the two dimensional polymerization of 1,3,5-trihydroxybenzene to GO.

\section*{Acknowledgements}

This work was supported in part by the Sofia University under research
grant No.70/05.04.2012. We thank an anonymous referee for his/her constructive comments which lead to a greatly improved presentation.


\end{document}